\documentclass[%
 reprint,
superscriptaddress,showkeys,showpacs,
 amsmath,amssymb,
 aps,pre
]{revtex4-2}

\usepackage{graphicx}
\usepackage{dcolumn}
\usepackage{bm}
\usepackage{amssymb}
\usepackage{bm}
\usepackage{xcolor}
\usepackage{appendix}
\usepackage{bigints}
\usepackage[utf8]{inputenc}
\usepackage{array}
\usepackage[caption=false]{subfig}

\newcommand{\be}{\begin{equation}}
\newcommand{\ee}{\end{equation}}
\newcommand{\ba}{\begin{array}}
\newcommand{\ea}{\end{array}}
\newcommand{\bqa}{\begin{eqnarray}}
\newcommand{\eqa}{\end{eqnarray}}
\newcommand{\bea}{\begin{eqnarray}}
\newcommand{\eea}{\end{eqnarray}}

\begin{document}


\title{Space Curves from Nonlinear Schr\"odinger Solutions: A Direct Approach}

\author{Kumar Abhinav}
\email{kumar.abh@mahidol.ac.th}
 \affiliation{Centre for Theoretical Physics and Natural Philosophy,\\
   Mahidol University, Nakhonsawan Campus, Phayuha Khiri, Nakhonsawan 60130, Thailand.}
\author{Partha Guha}%
 \email{partha.guha@ku.ac.ae}
\affiliation{Khalifa University of Science and Technology,\\
   PO Box 127788, Abu Dhabi, UAE.}
\date{\today}

\begin{abstract}
 The connection between vortex filament evolution in the local induction approximation and nonlinear Schr\"odinger (NLS) equation by Hasimoto [H. Hasimoto, J. Fluid Mechanics {\bf 51}, (1972) 477] has led to space curves corresponding to NLS solitons in the past. Utilizing this map, we propose a direct construction of parametric curve evolution from any NLS solution. It includes ordered (or nested) integrals of products of local matrices akin to the causal evolution of quantum theory, necessitating the implementation of the Magnus expansion. Such a straightforward mapping may be a simple tool to study the evolution of various systems of physical concern, although the actual computation can be a challenge for most NLS solutions.
\end{abstract}


\maketitle

Evolution vortex filaments in perfect fluids and classical one-dimensional continuum Heisenberg spin chain \cite{01} etc may mathematically be modeled as a moving space curve parameterized as $\gamma (s,t) \in {\mathbb R}^3$. Da Rios \cite{PG1} derived a set of two coupled equations governing the in-extensional motion of a vortex filament within an irrotational fluid in terms of time-evolving curvature
($\kappa(s,t)$) and torsion ($\tau(s,t)$) of the filament. 
\bea
&&\kappa_t+2\tau\kappa_s+\tau_s \kappa=0\nonumber\\
{\rm and}\quad&&\tau_t+\big(\tau^2-\frac{1}{2}\kappa^2-\frac{\kappa_{ss}}{\kappa} \big)_s=0.
\eea
These equations were rediscovered by Betchov decades later \cite{PG2}.

\paragraph*{}Hasimoto showed that these Da Rios–Betchov equations can elegantly be combined to give a focusing-type nonlinear Schr\"odinger (NLS) equation \cite{N2}, 
\be
q_t-iq_{ss}-2i\eta\vert q\vert^2q=0,\label{N4}
\ee
using the so-called Hasimoto map \cite{H4}:
\be
q(s,t)=\kappa(s,t)\exp\left\{i\int_0^s\tau(s',t)ds'\right\},\label{N5}
\ee
that relates $\tau(s,t)$ and $\kappa(s,t)$ to the NLS solution $q(s,t)$. Subsequently, the physical properties of the NLS system like densities of energy ($\kappa^2$) and momentum ($\kappa^2\tau$) can directly be obtained from these geometric parameters. In contrast, the non-linear coupling $\eta$ reflects the on-site interaction strength of the physical system. Up to a rigid motion, these equations prescribe the evolution of a vortex filament in an infinite domain of ${\mathbb R}^3$ for given initial conditions $\kappa(s,0)$ and $\tau (s,0)$. In particular, the single-soliton solution of this equation describes a helical twisting motion along the vortex line \cite{PG3}.

\paragraph*{}Extending Hasimoto's work, Lamb \cite{RNN01} studied the dynamics of space curves corresponding to soliton solutions of a class (sine-Gordon, mKdV, and NLS) of nonlinear differential equations. These methods primarily relied on exploiting the symmetries of the soliton structures to arrive at suitable ansatz solutions for Riccati-type differential equations. In addition to the Hasimoto map, other ways of connecting solitons to Frenet-Serret curves were obtained later for the same class of systems \cite{RNN02}. However, the construction of the space curve from an {\it arbitrary} solution of a given differential equation (the reconstruction problem) was still lacking \cite{RNN03,PG4}. The initial computation of an inverse Hasimoto transformation to reformulate the Da Rios–Betchov type equations had been performed by Sym \cite{PG4} and Aref-Linchem \cite{PG5}.

\paragraph*{}In this letter we attempt a direct inversion that leads to Frenet-Serret curve dynamics from the NLS solution, which has not been achieved previously to the best of our knowledge. 
It involves solving the Frenet-Serret equations in the matrix form that necessarily introduces $s$-integrals of ordered products of matrices containing curvature and torsion; quantities that can be read off of a given NLS solution. Unlike the reconstruction of space curves from particular soliton structures \cite{RNN01,H4,RNN02}, a general reconstruction approach necessarily requires either solving a third-order differential equation \cite{H4}, but with arbitrary local coefficients, or evaluating these exponentiated nested integrals of ordered local parameters. We carry out the latter task by invoking the Magnus expansion \cite{Magnus} that serves as a sensible estimate \cite{Mag01}. This Lie-algebra based approach has widely been used to determine curve dynamics representing the Cosserat rod motion \cite{SoftRob1} that models a soft manipulator in robotics \cite{SoftRob2}. 

\paragraph*{}Due to the computational complications, the accuracy of the present approach critically depends on the particular NLS solution. We demonstrate the procedure for two different NLS solutions by going up to the 4th order of the expansion. Beyond the formal interest, this approach may directly probe the dynamics and stability of various physical systems ({\it e.g.} spin configuration, fluids flows, etc.), which are effectively represented through NLS solutions, through the curve dynamics. Such correspondence between physical systems and NLS systems \cite{Shiva,NN1} with subsequent deformations \cite{OwnEPJB} have been studied before. 

\paragraph*{}The Frenet-Serret equations, that govern the vortex filament dynamics, can be cast into
the matrix form \cite{Rehan}:

\be
\boldsymbol{W}_s={\cal A}(s,t)\boldsymbol{W},\label{14}
\ee
where $\boldsymbol{W}=\left[{\bf t}~{\bf n}~{\bf b}\right]^T$ and 

\be
{\cal A}(s,t)=\left[ {\begin{array}{ccc}
   0 & \kappa(s,t) & 0 \\
   -\kappa(s,t) & 0 & \tau(s,t) \\
   0 & -\tau(s,t) & 0 \\
  \end{array} } \right].\label{15}
\ee
Here the parametric space derivatives of the tangent ($\boldsymbol{t}$), normal ($\boldsymbol{n}$) and binormal ($\boldsymbol{b}$) of a space-time curve are interrelated through corresponding $\kappa$ and $\tau$.
As the Hasimoto map relates the curvature and torsion to the NLS amplitude $\psi(s,t)$, the IHM amounts to solve for $\boldsymbol{W}$ that in turn should yield the tangent $\boldsymbol{t}=\boldsymbol{x}_s$. Then, on integrating the components, the three-dimensional coordinates $\{x_i\}$ of the curve can be obtained as a function of $s$ and $t$.   
\begin{figure}[bt]
    \centering
    \begin{minipage}{0.24\textwidth}
        \centering
        \includegraphics[width=\linewidth]{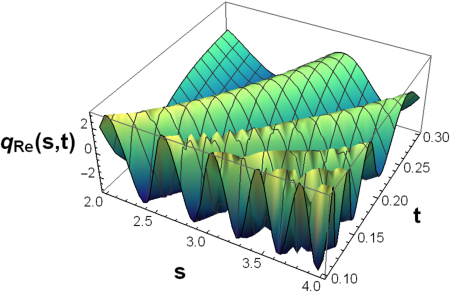}
        (a) Real part.
    \end{minipage}%
    \begin{minipage}{0.24\textwidth}
        \centering
        \includegraphics[width=\linewidth]{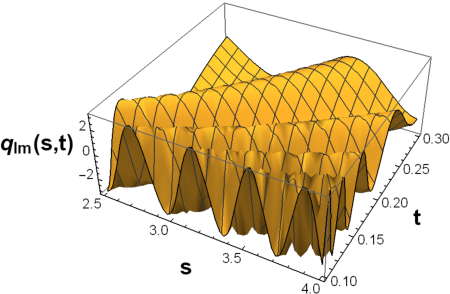}        
        (b) Imaginary part.
    \end{minipage}
    \caption{Plot for the periodic solution of the NLSE.}
    \label{Fig1}
\end{figure}
\paragraph*{} Although of first-order, Eq. \ref{14} is matrix-valued with local parameters and thus cannot be integrated. Solving individual equations by techniques subjected to particular (solitonic) $q(s,t)$ has been the approach in the past \cite{RNN01,H4,RNN02}. We notice that the general situation is identical to the time evolution of a quantum state under a time-dependent Hamiltonian invoking time-ordering of the unitary evolution \cite{Sakurai}. In the same way, respecting the matrix order, the general solution of Eq. \ref{14} takes the form:

\be
\boldsymbol{W}(s,t)=\mathfrak{S}\left(\exp\left[\int_0^s{\cal A}(s',t)ds'\right]\right)\boldsymbol{W}_0(t),\label{16}
\ee
with $\boldsymbol{W}_0(t)=\boldsymbol{W}(s=0,t)$. The symbol $\mathfrak{S}$ marks ordered product:

\bea
&&\mathfrak{S}\left({\cal A}(s_1,t){\cal A}(s_2,t)\right)\nonumber\\
&&:=\Theta(s_1-s_2){\cal A}(s_1,t){\cal A}(s_2,t)+\Theta(s_2-s_1){\cal A}(s_2,t){\cal A}(s_1,t);\nonumber\\
&&\text{where}\quad\Theta(s_1-s_2)=\begin{cases}
               1~\text{for}~s_1>s_2\\
               0~\text{otherwise}
            \end{cases},\label{NE1}
\eea
among the products of the integrands in the exponent. These integrals of ordered products can equivalently be written as {\it nested integrals} of the form,

\bea
&&\frac{1}{n!}\mathfrak{S}\left(\int_0^sds_1{\cal A}_1\int_0^sds_2{\cal A}_2\cdots\int_0^sds_n{\cal A}_n\right)\nonumber\\
&&=\int_0^sds_1{\cal A}_1\int_0^{s_1}ds_2{\cal A}_2\cdots\int_0^{s_{n-1}}ds_n{\cal A}_n,\label{NE2}\\
&&{\rm where}\quad{\cal A}_j:={\cal A}(s_j,t).\nonumber
\eea
Such results appear often in physical systems, tempting possible interpretations of the vortex filament evolution, more so as the matrix ${\cal A}$ is anti-Hermitian making the ordered exponent in Eq. \ref{16} unitary.
\begin{figure}[bt]
    \begin{minipage}{0.24\textwidth}
        \centering
        \includegraphics[width=\linewidth]{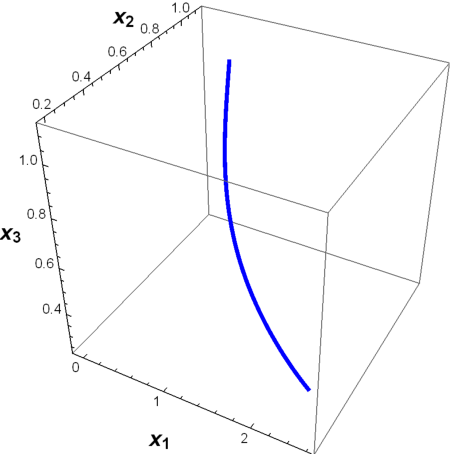}
        (a) $t=2$
    \end{minipage}%
    \begin{minipage}{0.24\textwidth}
        \centering
        \includegraphics[width=\linewidth]{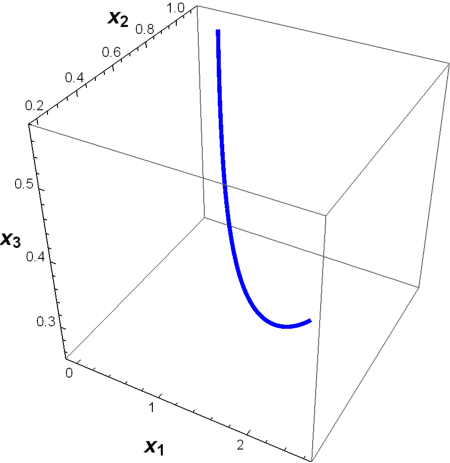}        
        (b) $t=4$
    \end{minipage}
    \vskip\baselineskip
    \begin{minipage}{0.24\textwidth}
        \centering
        \includegraphics[width=\linewidth]{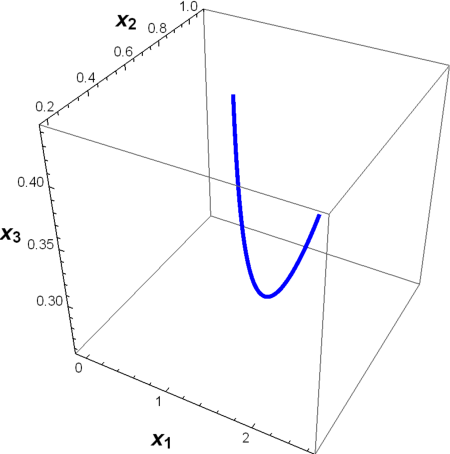}
        (c) $t=6$
    \end{minipage}%
    \begin{minipage}{0.24\textwidth}
        \centering
        \includegraphics[width=\linewidth]{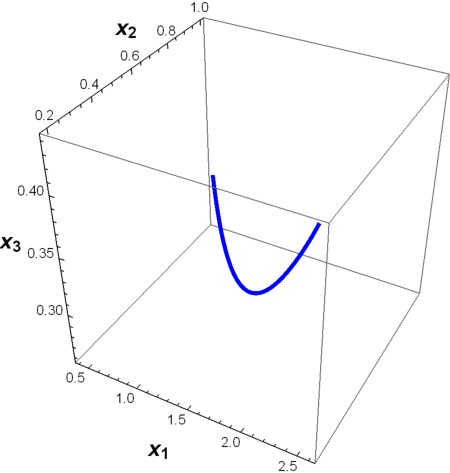}        
        (d) $t=8$
    \end{minipage}
    \caption{Frenet-Serret curve for the periodic NLS solution with time $t$ as a parameter.}
    \label{Fig2}
\end{figure}
\paragraph*{}The exponentiated nested integrals in $\boldsymbol{W}(s,t)$ can be be estimated through the Magnus expansion \cite{Magnus} in the form:
\bea
&&\boldsymbol{W}(s,t)=\boldsymbol{W}(0,t)\exp\boldsymbol{\Omega}(s,t),\quad\boldsymbol{W}(0,t)=\mathbb{I},\nonumber\\
&&\boldsymbol{\Omega}(s,t)=\sum_{k=1}^\infty\boldsymbol{\Omega}_k(s,t).\label{NEm1}
\eea 
The choice of $\boldsymbol{W}(0,t)$ is arbitrary and so it can be chosen as the identity matrix for brevity. The first few terms of the expansion are,
\begin{widetext}
\bea
&&\boldsymbol{\Omega}_1(s,t)=\int_0^sds_1\,{\cal A}_1,\nonumber\\
&&\boldsymbol{\Omega}_2(s,t)=\frac{1}{2}\int_0^sds_1\int_0^{s_1}ds_2\,\left[{\cal A}_1,\,{\cal A}_2\right],\nonumber\\
&&\boldsymbol{\Omega}_3(s,t)=\frac{1}{6}\int_0^sds_1\int_0^{s_1}ds_2\int_0^{s_2}ds_3\,\Bigg(\left[{\cal A}_1,\,\left[{\cal A}_2,\,{\cal A}_3\right]\right]+\left[\left[{\cal A}_1,\,{\cal A}_2\right],\,{\cal A}_3\right]\Bigg),\nonumber\\
&&\boldsymbol{\Omega}_4(s,t)=\frac{1}{12}\int_0^sds_1\int_0^{s_1}ds_2\int_0^{s_2}ds_3\int_0^{s_3}ds_4\,\Bigg(\left[\left[\left[{\cal A}_1,\,{\cal A}_2\right],\,{\cal A}_3\right],\,{\cal A}_4\right]+\left[{\cal A}_1,\,\left[\left[{\cal A}_2,\,{\cal A}_3\right],\,{\cal A}_4\right]\right]\nonumber\\
&&\qquad\qquad\qquad\qquad\qquad+\left[{\cal A}_1,\,\left[{\cal A}_2,\,\left[{\cal A}_3,\,{\cal A}_4\right]\right]\right]+\left[{\cal A}_2,\,\left[{\cal A}_3,\,\left[{\cal A}_4,\,{\cal A}_1\right]\right]\right]\Bigg),\nonumber\\
&&\vdots\label{NEm2}
\eea
\end{widetext}
The Magnus expansion up to the 4th order is considered to be a sensible estimate of the ordered exponentiated integral \cite{Mag01}, leading to a viable estimation for $\boldsymbol{W}$.
Since the first row of the matrix $\boldsymbol{W}$ contains the components of the tangent, the identification $dx_i/ds=W_{1i}(s,t)$ then immediately provides the space coordinates of the moving curve,

\bea
x^i(s,t)=\int_0^sW_{1i}(s',t)ds'+x^i_0(t),\label{20}
\eea
subjected to arbitrary initial conditions $x^i_0(t)$. 

\paragraph*{}From Eq. \ref{N5}, given a particular NLS solution $q(s,t)$ the curvature and torsion are expressed in terms of the NLS amplitude and phase as,
\bea
&&\kappa(s,t)=\vert q(s,t)\vert\nonumber\\
{\rm and}\quad&&\tau(s,t)=-i\frac{d}{ds}\log\left(\frac{q(s,t)}{\vert q(s,t)\vert}\right).\label{23}
\eea
\begin{figure}[tt]
    \centering
    \begin{minipage}{0.24\textwidth}
        \centering
        \includegraphics[width=\linewidth]{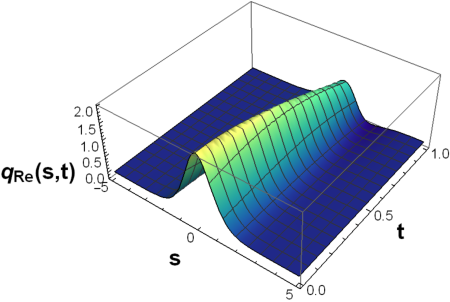}
        (a) Real part.
    \end{minipage}%
    \begin{minipage}{0.24\textwidth}
        \centering
        \includegraphics[width=\linewidth]{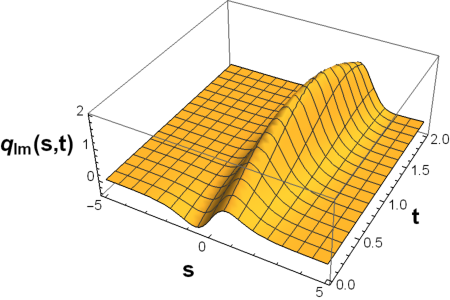}        
        (b) Imaginary part.
    \end{minipage}
    \caption{Plot for the soliton solution of the NLSE.}
    \label{Fig3}
\end{figure}
A case of particular interest is that of the focusing-type NLS system of Eq. \ref{N4} related to the spin orientation of the 1-D Heisenberg XXX model in the continuum limit \cite{N1}. The spin orientation itself is identified with the tangent to the Frenet-Serret curve, implying a direct physical interpretation of the curve dynamics. Since any NLS solution, not just solitonic ones, qualifies for the form in Eq. \ref{N5} we demonstrate the construction of the vortex filament dynamics for a particular periodic solution to the NLSE obtained by Sall' \cite{Sall}:

\be
q(s,t)=\frac{1}{\sqrt{t}}\exp i\left(\frac{s^2}{4t}+2\ln{t}\right),\label{SolZak}
\ee
plotted in Fig. \ref{Fig1}. The solution is well-behaved and quite amiable to the calculation of the matrix exponent, which is not the case for other NLS solutions as will be seen. 
Then the curvature and torsion take particular forms,
\be
\kappa(s,t)=\frac{1}{\sqrt{t}}\quad{\rm and}\quad\tau(s,t)=\frac{s}{2t},
\ee
on substituting which the Magnus exponent can be obtained as,
\begin{widetext}
\be
\boldsymbol{\Omega}(s,t)=\begin{pmatrix}
    0 & y-\frac{y^5}{960} & \frac{y^3}{24}+\frac{y^5}{1440}\left(1+\frac{3y^2}{56}\right)\\
    -y+\frac{y^5}{960} & 0 & \frac{y^2}{4}\\
    -\frac{y^3}{24}-\frac{y^5}{1440}\left(1+\frac{3y^2}{56}\right) & \frac{y^2}{4} & 0
\end{pmatrix},\quad y:=\frac{s}{\sqrt{t}}.\label{matrix1}
\ee
\end{widetext}
up to the 4th order. 

\paragraph*{}Following exponentiation of $\boldsymbol{\Omega}(s,t)$, the vortex filament obtained is plotted in Fig. \ref{Fig2} using {\it Mathematica} 12 that evolves in the $\{x_i\}$ space with the NLS time parameter $t$. Although a more vigorous treatment is needed for a clearer demonstration, the curvature at any point decreases with time so the torsion is as expected. Going beyond the 4th order should incorporate more features since the present range of validity is bound by the convergence condition of the Magnus expansion $\int_0^s\left\vert\left\vert{\cal A}(s',t)\right\vert\right\vert_2\leq\pi$ \cite{MagConv} that, for example, requires $s\lesssim 1.95$ for $t=4$.

\paragraph*{}Relatively complicated NLS solutions require numerical approximations as the corresponding Magnus integrals do not possess closed forms. A worthwhile demonstration of this can be with the NLS soliton (Fig. \ref{Fig3}),
\bea
q(s,t)&=&2\rho^2{\rm sech}^2\left[\rho\left(s-vt-s_0\right)\right]\nonumber\\
&\times&\exp\left[i\left(\rho^2t-\frac{v^2}{4}t+\frac{v}{2}s\right)\right],
\eea
with velocity $v$ and frequency $\rho v$. Given their wide-spread importance, various soliton solutions have widely been used \cite{RNN01,H4} for constructing space curves, more so as the corresponding Riccati equations obtained from the Frenet-Serret system possess analytical solutions \cite{RNN01}. For the above NLS soliton, the space curve corresponds to curvature and torsion,
\be 
\kappa(s,t)=2\rho^2{\rm sech}^2\left[\rho\left(s-vt-s_0\right)\right],\quad\tau(s,t)=v.
\ee
\begin{figure}[tt]
    \begin{minipage}{0.24\textwidth}
        \centering
        \includegraphics[width=\linewidth]{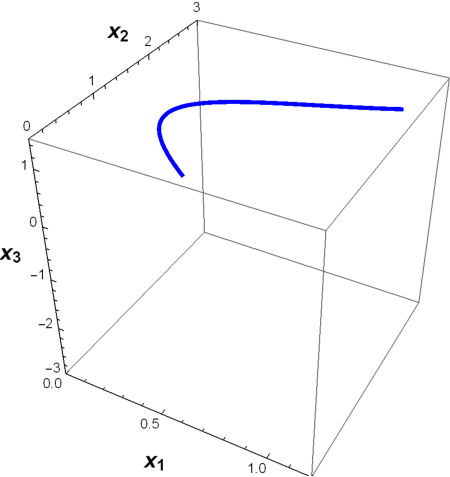}
        (a) $t=0$
    \end{minipage}%
    \begin{minipage}{0.24\textwidth}
        \centering
        \includegraphics[width=\linewidth]{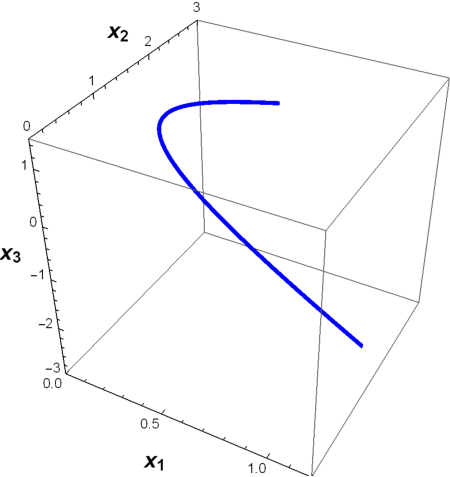}        
        (b) $t=1$
    \end{minipage}
    \vskip\baselineskip
    \begin{minipage}{0.24\textwidth}
        \centering
        \includegraphics[width=\linewidth]{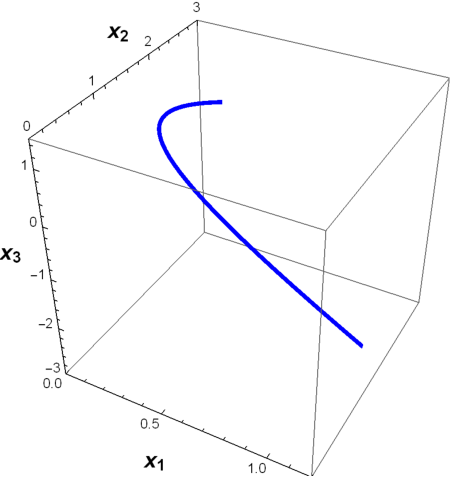}
        (c) $t=1.5$
    \end{minipage}%
    \begin{minipage}{0.24\textwidth}
        \centering
        \includegraphics[width=\linewidth]{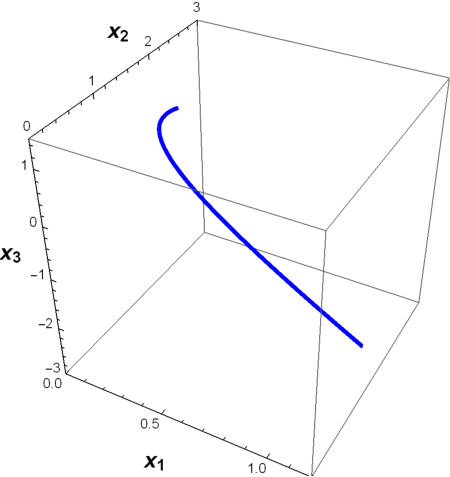}        
        (d) $t=2$
    \end{minipage}
    \caption{Evolution of the Frenet-Serret curve corresponding to the NLS soliton with $t$ as a parameter.}
    \label{Fig4}
\end{figure}

Not unlike Hasimoto's observation \cite{H4}, this constant torsion curve has a maximum curvature of $2\rho^2$ at $s=vt$ that falls off asymptotically. The result is a single twisted maximum whose projection normal to its motion is a closed loop \cite{H4,RNN01,PG3}. We choose $\rho=1=v$ and initial conditions to be $s_0=0$ for brevity. For this soliton, the present approach does not yield a closed-form beyond the 3rd order of the Magnus expansion. We employ numerical approximations to arrive at the Magnus exponent that retains the anti-symmetry as before, having non-vanishing elements,
\bea
&&\Omega_{12}=2 \tanh (z)+\frac{1}{6} (4 \log (\cosh (z))-2 z \tanh (z))\nonumber\\
&&\qquad=-\Omega_{21},\nonumber\\
&&\Omega_{13}=\frac{1}{6} \Bigg[-3 \text{Li}_3\left(-e^{-2 z}\right)-z \Big\{3 \text{Li}_2\left(-e^{-2 z}\right)\nonumber\\
&&\qquad+8 \tanh (z)\Big\}-z^3+z^2 \log (\cosh (z))+4 \text{sech}^2(z) \nonumber\\
&&\qquad\times\Big\{\cosh (2 z) \log (\cosh (z))-2\Big\}\Bigg]\nonumber\\
&&\qquad+\frac{1}{2} \left[2 z \tanh (z)-4 \log (\cosh (z))\right]=-\Omega_{31},\nonumber\\
&&\Omega_{23}=z+\frac{1}{6} \Big(-12 z+12 \tanh (z)+2 x \text{sech}^2(z)\nonumber\\
&&\qquad+12 \tanh (z) \log (\cosh (z))\Big)=-\Omega_{32},\\
&&\text{where}~z=s-t.\nonumber
\eea
The vortex filament evolution is obtained numerically as shown in Fig. \ref{Fig4}. Though the full helical structure is not clear, the curve expectantly traverses a region with maximum curvature having visibly uniform torsion. A more extensive computational approach should yield a localized helical curve which is beyond the scope of the present qualitative proposal.

\paragraph*{}In conclusion, we have obtained the vortex filament evolution corresponding to general NLS solutions by directly solving the Frenet-Serret set of equations. The resultant ordered integrals contain products of curve parameters derived from the amplitude and phase of the particular solutions. The Magnus expansion has been employed to estimate these integrals, leading to closed-form expressions in the case of the periodic NLS solution. However, the Magnus expansion with parameters from the soliton solution requires numerical approximations to arrive at the usual 4th order of expansion. Beyond the present introductory qualitative demonstration, extensive numerical tools should lead to more elaborated and quantitatively distinct space curves corresponding to different NLS solutions. In a more practical sense a wider understanding of physical systems, wherein Hasimoto map has been useful recently for geometrical understanding of the soliton sector only {\it e.g.} the Landau-Lifshitz system that encompasses Heisenberg ferromagnetism \cite{K1} and optical fluids \cite{K2}, can become possible. For more arbitrary solutions the present utilization of the Magnus expansion may enable a wider physical implications of NLS and similar nonlinear systems, including the aforementioned area of soft robotics \cite{SoftRob1,SoftRob2}

\vskip 0.5cm
\noindent{\it Acknowledgement:} Kumar Abhinav's work enjoys the support of Mahidol University. The work of Partha Guha is supported by the Khalifa University of Science and Technology, United Arab Emirates under Grant Number FSU-2021-014.

\end{document}